\documentclass[journal]{IEEEtran}

\usepackage{cite}
\usepackage{amsmath,amssymb,amsfonts,bm}
\usepackage{algorithmic}
\usepackage{graphicx}
\usepackage{textcomp}
\usepackage{xcolor}
\usepackage{multicol}
\usepackage{booktabs}
\usepackage{multirow}
\usepackage{gensymb}

\usepackage{soul}
\usepackage{graphicx}
\usepackage{amsmath}
\usepackage{booktabs}
\usepackage{threeparttable}
\usepackage[hidelinks]{hyperref}

\newtheorem{definition}{Definition}[section]
\newtheorem{observation}{Observation}[section]

\newcommand*{\vect}[1]{\text{\textbf{#1}}}

\newcommand*{\flr}[1]{\left\lfloor #1 \right\rfloor}

\newcommand*{\temp}[1]{#1~{\degree}C}

\begin{document}

\title{Enhancing Strong PUF Security with Non-monotonic Response Quantization}

\author{\IEEEauthorblockN{Kleber Stangherlin, Zhuanhao Wu, Hiren Patel, Manoj Sachdev} \\
        \IEEEauthorblockA{ECE Department, University of Waterloo, Waterloo, ON N2L 3G1, Canada}\\
        \{khstangh, zhuanhao.wu, hiren.patel, msachdev\}@uwaterloo.ca
}

\markboth{Submitted to IEEE for possible publication. Copyright may be transferred. This version may no longer be accessible.}{}

\maketitle

\begin{abstract}
Strong physical unclonable functions (PUFs) provide a low-cost authentication primitive for resource constrained devices. However, most strong PUF architectures can be modeled through learning algorithms with a limited number of CRPs. In this paper, we introduce the concept of non-monotonic response quantization for strong PUFs. Responses depend not only on which path is faster, but also on the distance between the arriving signals. Our experiments show that the resulting PUF has increased security against learning attacks. To demonstrate, we designed and implemented a non-monotonically quantized ring-oscillator based PUF in 65~nm technology. Measurement results show nearly ideal uniformity and uniqueness, with bit error rate of 13.4\% over the temperature range from \temp{0} to \temp{50}.
\end{abstract}


\section{Introduction}

Strong physical unclonable functions (PUFs) provide a low-cost authentication primitive for resource constrained devices. PUFs harvest process variability to generate a fingerprint of an integrated circuit (IC). Authentication is performed with an externally accessible, unencrypted, challenge-response protocol. Fabricating two identical PUFs is unfeasible even for the original manufacturer, making it a promising solution for low-cost authentication \cite{seminalArb2002}. Ideally, strong PUFs do not leak information about its internal characteristics. In real implementations, however, it's been shown that PUF responses carry significant information about its internal entropy source. Using a limited number of CRPs, learning attacks are capable of predicting strong PUF responses for unseen challenges \cite{seminalAttack2013}.

\begin{figure}[t]
    \centering
    \includegraphics[scale=1]{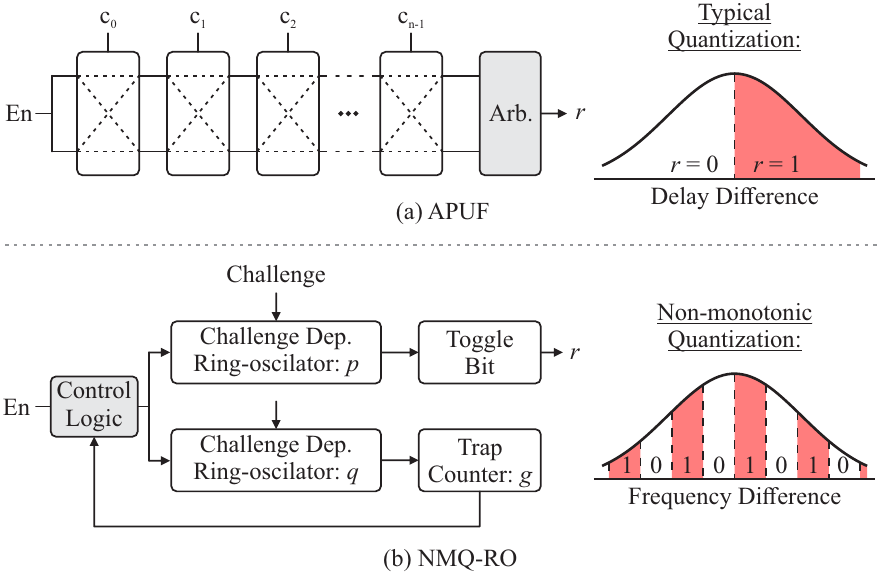}
    \caption{Arbiter PUF with typically quantized response in (a); and ring-oscillator based PUF with non-monotonically quantized response in (b).}
    \label{fig:arch}
\end{figure}

Many strong PUF architectures, at their core, encode a comparison result of two identical structures. For example, Fig. \ref{fig:arch} (a) illustrates the popular arbiter PUF (APUF) with two identical delay lines \cite{seminalArb2002}. Each $n$-bit challenge performs a unique selection of delay elements for the two paths. Depending on which path is faster, the arbiter makes a binary (quantized) decision. The APUF uses \textit{typical response quantization}, where the decision remains the same regardless of the distance between the arriving signals. It's been shown that APUF, and many of its variants, can be modeled using learning algorithms \cite{seminalAttack2013, attackRelb2015, attackIPUF2021, attackIPUFSplit2020}. In this work, we innovate by changing the core quantization technique which is common to most previous strong PUF designs. We introduce the concept of \textit{non-monotonic response quantization} (NMQ) to increase the security of strong PUF architectures. The quantized decision depends not only on which path is faster, but also on the distance between the arriving signals.

A non-monotonically quantized strong PUF (NMQ-PUF) can take various forms. We demonstrate the technique using a ring-oscillator based architecture denoted as NMQ-RO, shown in Fig. \ref{fig:arch} (b). It uses two challenge dependent oscillators. One oscillator is connected to a counter, and the other to a toggling bit. The control logic allows oscillators to run until the counter reaches a predefined value. The response is taken from the toggling bit. The final counter value is chosen such that responses, when plotted along the frequency difference axis, exhibit an alternating pattern of zeros and ones.

Our main contribution is the proposal of non-monotonic quantization for PUFs, which can be applied to a vast number of existing PUF architectures. Other key ideas introduced in this paper include the concept of uniqueness sensitivity to entropy source, which offers rapid insight on PUF resilience to learning attacks. Using measurement data from a fabricated testchip in 65~nm CMOS, we perform extensive experiments to explore the security/stability trade-off of our NMQ-PUF implementation. In particular, we show that a single NMQ-RO instance is resilient to logistic regression attacks, while compositions using multiple NMQ-ROs, denoted as $k$-XOR-NMQ-RO, are resistant to attacks using Fourier analysis, and deep neural networks trained with 20 M CRPs. Measured bit error rate for compositions using 2, and 3 NMQ-PUFs is less than 13.4\%, and 18.6\%, over the \temp{0} -- \temp{50} temperature range.

This paper is organized as follows. First, section \ref{sec:back} discusses the background, strong PUF performance metrics, and performs an analysis of the practical consequences of high bit error rate. Section \ref{sec:relworks} brings a review of relevant works, and current state-of-the-art in strong PUF design. Section \ref{sec:nmq} provides a definition for NMQ, along with a geometric interpretation of it. The details of our NMQ-PUF implementation are discussed in section \ref{sec:nmqro}. Testchip design and measurement results are reported in section \ref{sec:meas}. Section \ref{sec:sec} introduces the concept of uniqueness sensitivity to entropy source, and reports data obtained from our learning attack experiments. Section \ref{sec:conc} presents the conclusion.

\section{Background\label{sec:back}}

\begin{figure}[t]
    \centering
    \includegraphics[scale=1]{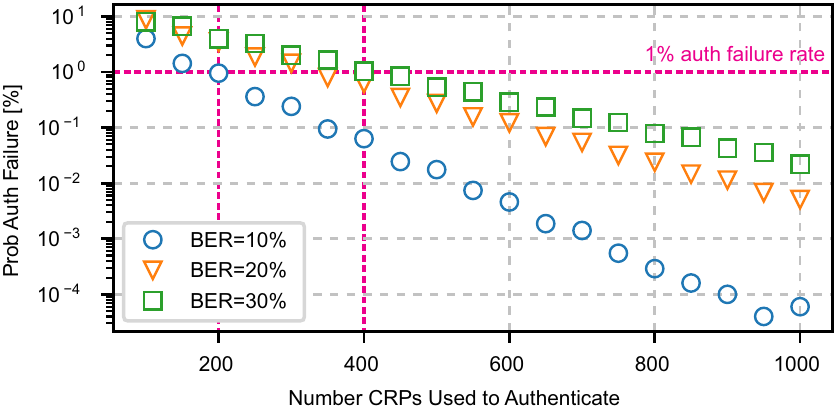}
    \caption{Probability of authentication failure simulated with 1M authentications. Uniformity of 50\% is assumed. The minimum number of correct responses is 5\% below (100\% - BER).}
    \label{fig:auth}
\end{figure}

The key difference of PUFs to programmed identifiers is the lack of blank samples. PUFs give ICs the capability of unique responses, no matter how their memory is programmed. Practical use of strong PUFs still requires the programming of a public chip identifier (ID) which is used to associate an enrolled CRP database to a particular PUF instance \cite{seminalArb2002}. Later, when performing an authentication, the programmed ID is read and the correct CRP database is fetched. The PUF is then inquired with a subset of the challenges in the database. Challenges are never used more than once. If the number of correct responses exceeds an application defined threshold, the IC is deemed authentic.

\subsection{Strong PUF Metrics}

Quality assessment of strong PUFs uses metrics that evaluate uniformity, uniqueness, and stability of responses.

\subsubsection{Uniformity} estimates the ratio of zeros and ones in PUF responses. It is also known as \textit{normalized hamming weight}. Ideal uniformity is 0.5, which indicates, on average, equal number of zeros and ones.

\subsubsection{Uniqueness\label{sec:bg:uniq}} estimates the distance between responses from multiple instances. It is also known as \textit{normalized hamming distance}. Ideal uniqueness is 0.5, which indicates that, for the same set of challenges, on average, half responses will differ.

\subsubsection{Bit error rate (BER)} estimates reproducibility of responses under several environmental conditions. Bit error rate (BER) reports a ratio of bits (responses) that differ from their enrolled value. BER ideal value is 0\%, which indicates no incorrect responses during measurement. Other literature may use the term reliability, which simply denotes (100\% - BER).

\subsection{Authenticating with Non-ideal BER}

To authenticate strong PUFs, the number of correct responses must exceed a response threshold, otherwise the authentication fails. The response threshold is a system defined parameter which is set according to PUF response stability. If the threshold is expressed as a percentage of correct responses needed to authenticate, we can argue that it must be less than (100\% - BER), otherwise the PUF is unlikely to successfully authenticate. 

Fig. \ref{fig:auth} simulates PUFs with different BER. Threshold was set 5\% below (100\% - BER). For example, when simulating an authentication using 200 CRPs, with a PUF that has 10\% BER, 170 correct responses are required to authenticate. As shown in Fig. \ref{fig:auth}, the probability of authentication failure falls exponentially with the number of CRPs used to authenticate. For example, for a 1\% failure rate, a PUF with 10\% BER will require 200 CRPs, while if the PUF BER is increased to 20\% or 30\%, the required CRPs to achieve the same failure rate will increase to 350, and 400, respectively.

\begin{figure*}[t]
    \centering
    \includegraphics[scale=1]{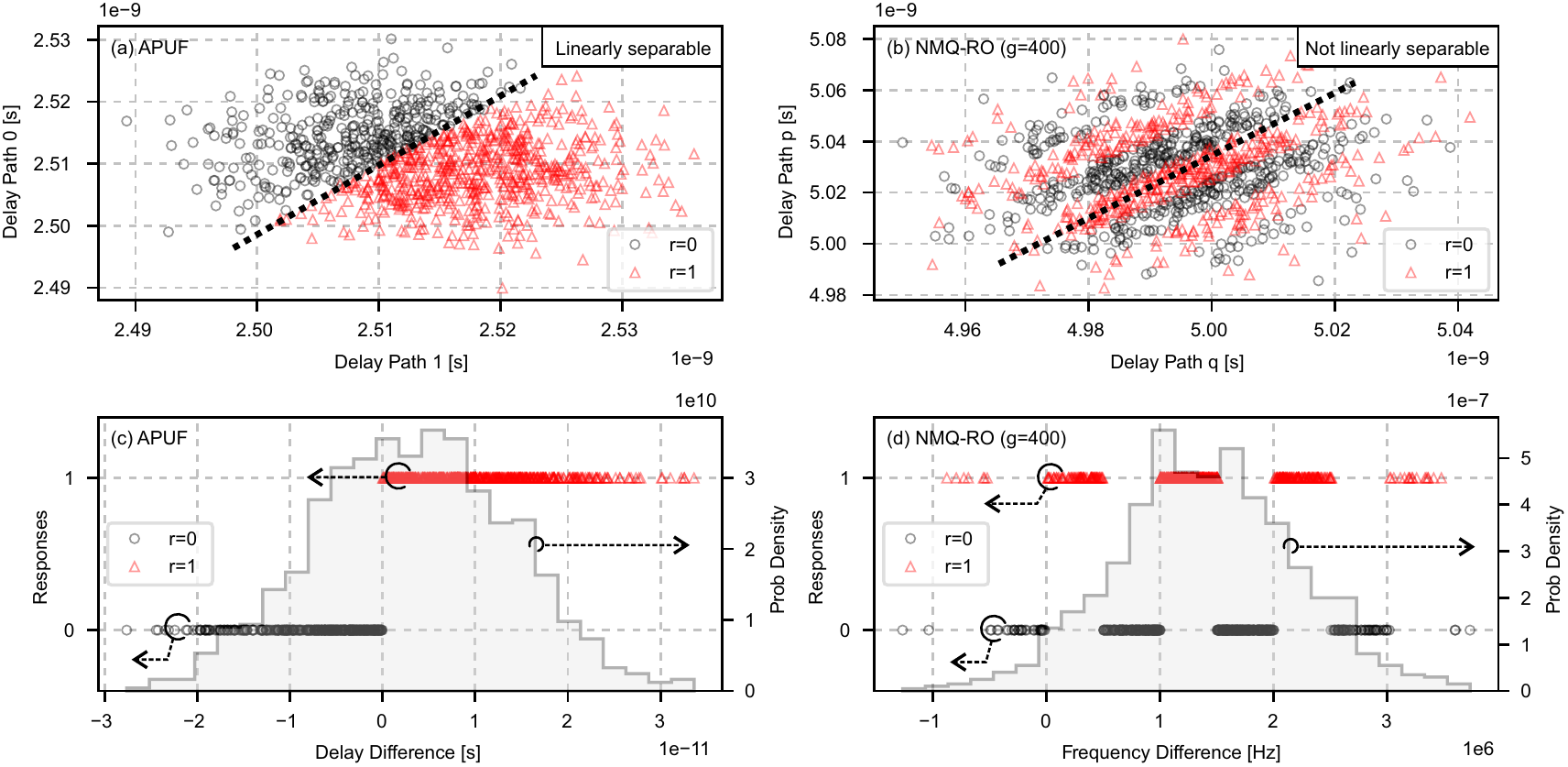}
    \caption{SPICE simulated responses for 1000 challenges; (a), and (b) plot APUF, and NMQ-RO (g=400) responses using accumulated path delay as basis, respectively; (c) and (d) plot APUF, and NMQ-RO (g=400) responses along the quantizer input axis, which is delay difference for APUF, and frequency difference for NMQ-RO; (c) and (d) also plot the delay and frequency difference histogram for APUF and NMQ-RO.}
    \label{fig:ovf}
\end{figure*}

\section{Related Works\label{sec:relworks}}

The arbiter PUF (APUF) was introduced as the first silicon PUF \cite{seminalArb2002}. PUF architectures have been continuously enhanced to improve resilience against learning attacks. The feed-forward PUF inserts extra arbiters in the delay path to reduce response linearity \cite{pufFeedforward2004}. Similarly, XOR-PUF used XOR gates to combine multiple APUFs \cite{pufXorAndROPool2007}, while lightweight PUF added input challenge transformation and a parity based output function \cite{pufLightweight2008}. Double arbiter PUF combines two APUFs using a different wiring method and an XOR gate \cite{pufDoubleArb2014}. The interpose PUF uses a composite architecture of two XOR-APUFs to generate a response \cite{pufInterpose2019}. Recent advances in PUF architectures require larger CRP datasets, and longer training time, but none was shown secure against learning attacks \cite{seminalAttack2013, attackRelb2015, attackDNN2019, attackIPUF2021, attackIPUFSplit2020}. Adding architectural, or wiring complexity that always encode the \textit{fastest path} has limited benefits, and may ruin response stability. We innovate by changing the core quantization technique, which is common to all above mentioned PUF designs.

Authors in \cite{seminalArb2002} discuss self-oscillating circuits and frequency comparison as a method for measuring on-chip delay, but no implementation details were presented. In \cite{pufXorAndROPool2007}, a strong PUF was implemented using a pool of ring-oscillators. Responses were obtained by pairwise frequency comparisons, which resulted in a limited challenge space. The architecture was attacked in \cite{seminalAttack2013}. Our work does not use pairwise frequency comparisons. We generate responses using non-monotonic quantization instead of directly encoding oscillator performance information.

In \cite{pufROMap2011}, authors proposed an RO PUF with an identity-mapping function to expand the set of CRPs. They also used error correction to reduce bit error rate. Their architecture was attacked in \cite{attackROMap2015} using vulnerabilities in the identity mapping function. Moreover, the use of error correction in strong PUFs was shown vulnerable in \cite{pufEccVulnerable2014} due to helper data manipulation. We use challenge dependent delay paths as entropy source. Our proposed implementation of NMQ has challenge space of $2^n$, where $n$ is 64. Our architecture allows designers to find a compromise between learning resilience and response stability, therefore, we do not employ error correction techniques.

Recent works have tried to decrease the bit error rate of PUFs. In \cite{pufDuty2016, pufDuty2018}, authors discuss the use of asymmetric ROs that implement duty-cycle comparison instead of frequency comparison, arguing that duty-cycles are less sensitive to temperature variations. In \cite{pufThyristor2016}, thyristor based delay cells are used for its improved temperature stability. In \cite{pufPolysi2016}, the use of poly-Silicon is discussed, where random grain boundaries and trapped charges offer larger process variation. However, none of these works investigates security of PUFs against learning attacks. Moreover, our implementation does not use any circuit techniques to improve response stability. Therefore, we can potentially benefit from the above mentioned methods to enhance temperature stability of our RO-based design.

In \cite{pufChaotic2021}, authors propose a strong PUF based on bistable ROs. Such oscillators use an even number of challenge dependent delay stages, acting as a large SRAM cell. After start, the values in each stage will eventually settle to either $01010 \ldots 01$, or $10101 \ldots 10$. To defend against learning attacks, authors added challenge dependent post-processing of responses using a chaotic algorithm. The security of the proposed system relies on the secrecy of implementation parameters used in the obfuscation algorithm. Our work uses a white-box approach for security evaluation. We assume the attacker has complete knowledge of the design. Moreover, our design avoids the extra area cost needed to implement obfuscation algorithms.

\section{Non-monotonic Quantization\label{sec:nmq}}

\subsection{Key Observations}

PUFs compare the manufacturing variability of identically designed circuit components. The input challenge specifies the structures to be compared. Responses always encode the comparison result, leaking valuable information about the \textit{best performing} circuit. For example, APUF responses immediately expose which delay path is faster, for a certain challenge. Other APUF-based compositions introduce additional instances in an attempt to obfuscate the entropy source information from the final responses, but they have generally failed to produce learning resistant strong PUFs, since the core quantization technique is still the same. Non-monotonic quantization (NMQ), however, produces outputs that can not be individually translated into information about the entropy source.

\medbreak
\begin{observation}
Typically quantized strong PUFs, at their core, encode the best performing structure. Individual CRPs provide direct information about the entropy source.
\end{observation}
\medbreak

A geometrical perspective of non-monotonic quantization is shown in Fig. \ref{fig:ovf}. In (a), APUF responses for 1000 challenges are plotted using accumulated path delay as basis. Similarly, (b) plots NMQ-RO responses. It is clear that the typically quantized responses in (a) are separable by a line. The NMQ responses in (b) however, can not be linearly separable on the chosen basis.

\medbreak
\begin{observation}\label{obs:linsep}
Non-monotonically quantized strong PUF responses are not linearly separable on the basis of accumulated path delay.
\end{observation}
\medbreak

Based on Obs. \ref{obs:linsep}, we may argue that linear models, such as logistic regression (LR), are not able to model NMQ-PUFs. While APUFs require composition to avoid LR attacks, our NMQ-PUF implementation is shown resistant against LR, even when responses are taken from a single instance. Detailed experimental data is presented in section \ref{sec:sec}.

Fig. \ref{fig:ovf} (c) and (d) plot APUF and NMQ-RO responses along the delay, and frequency difference axis, respectively. While typically quantized responses from the APUF encode the delay difference sign, NMQ-RO responses show an alternating pattern of zeros and ones. The delay and frequency difference histograms are plotted above responses in (c) and (d), for APUF, and NMQ-RO. Unlike APUF, the NMQ-RO histogram is not necessarily centered at zero. NMQ-RO responses are defined by the distance between the two frequencies, not the sign of their difference.

\section{NMQ-PUF Using Ring-oscillators\label{sec:nmqro}}

A non-monotonically quantized strong PUF (NMQ-PUF) can take various forms. We demonstrate a ring-oscillator based strong PUF, denoted NMQ-RO, which is shown in Fig. \ref{fig:arch}. It uses two identical, challenge dependent ring-oscillator structures. One ring-oscillator is connected to a counter named \textit{trap counter}, and the other to a \textit{toggling bit}. The toggling bit is complemented, and the trap counter is incremented, at every rising edge of their associated oscillating signal. The control logic disables both ring oscillators when the trap counter reaches a predetermined value. The final response is taken from the toggling bit.

The response, $r$, may be written in terms of circuit variables as

\begin{equation}
r = \text{LSB}\left(\flr{g\frac{D_p(\vect{c})}{D_q(\vect{c})}}\right), \label{eq:r}
\end{equation}

where $\text{D}_{\{p,q\}}(\vect{c})$ are the challenge dependent propagation delays of ring oscillator $p$, and $q$. The term $g$ denotes the pre-determined trap counter final value. The function $\text{LSB}()$ returns the least significant bit.

According to Eq. \ref{eq:r}, NMQ-RO responses encode the performance ratio of two ring-oscillators into a single bit value that can only represent $\{0, 1\}$. Non-monotonic quantization arises due to information lost during the encoding of $D_p(\vect{c})/D_q(\vect{c})$ into the single bit response $r$. For example, if

\begin{equation*}
g\frac{D_p(\vect{c})}{D_q(\vect{c})} \geq 2,
\end{equation*}

the performance ratio information, for that particular challenge, is no longer directly encoded in the response, since only the least significant bit of the above product is captured. From a design perspective, we can promote information loss by increasing the trap counter final value, $g$. The parameter $g$ must be selected such that information loss occurs for a significant number of challenges.

Large $g$ values, however, increase bit error rate. Our measurement results and security assessment, sections \ref{sec:meas} and \ref{sec:sec}, indicate that large $g$ enhances learning resilience, but impacts response stability. Designers should explore the trade-off between security and response stability, choosing the parameter $g$ accordingly.

\section{Compositions of NMQ-PUFs}

The security of NMQ-RO is primarily controlled by the chosen $g$ value. However, our measurement data reveals that $g$ sets a trade off between learning attack resilience and response stability. As shown in section \ref{sec:sec}, improving security solely by increasing the $g$ can have a significant impact on bit error rate.

In order to find a compromise between security and response stability, we explore PUF compositions. Researchers have demonstrated that compositions enhance APUF resilience against learning attacks \cite{seminalAttack2013}. Therefore, we apply similar methods to NMQ-RO, but instead of using typically quantized PUFs, we use NMQ-RO with a moderate final trap counter value (g=200), such that BER for the overall composition is less than 20\%.

\begin{figure}[t]
    \centering
    \includegraphics[scale=1]{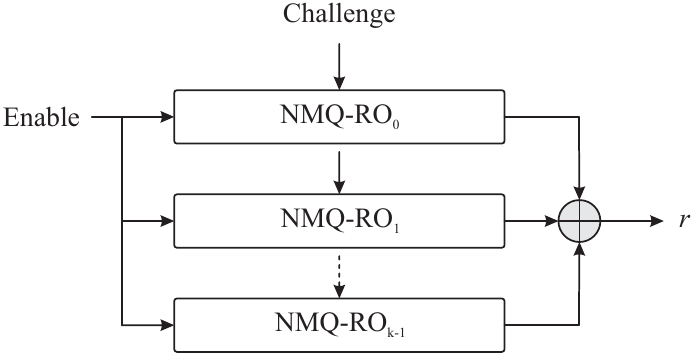}
    \caption{XOR-NMQ-RO composition. Same challenge is applied to all instances.}
    \label{fig:xorpuf}
\end{figure}

We demonstrate a composite strong PUF using three NMQ-ROs. The $k$-XOR-NMQ-RO is shown in Fig. \ref{fig:xorpuf}. Term $k$ refers to the number of NMQ-RO instances used. Similarly to a $k$-XOR-APUF, all instances evaluate the same challenge. Outputs are XORed to produce the final response. 

Other PUF compositions could also increase the resilience to learning attacks, while maintaining response stability at an acceptable level. We chose the XOR-APUF style of composition because it is well understood, with mature attack techniques.

\begin{figure}[t]
    \centering
    \includegraphics[scale=1]{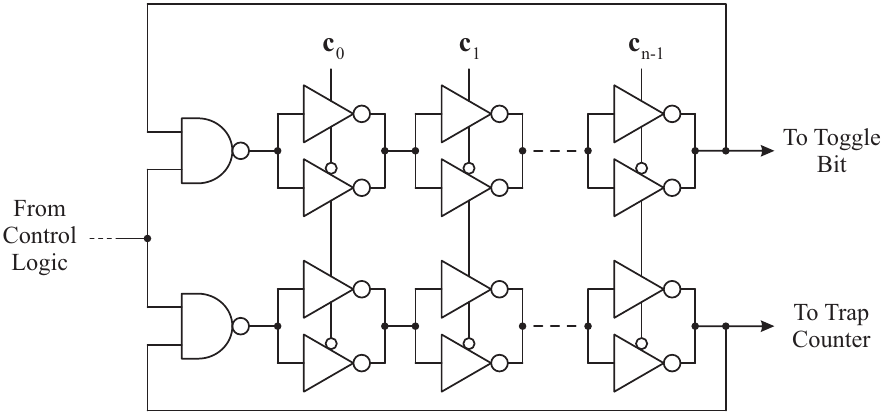}
    \caption{CMOS implementation of the NMQ-RO challenge dependent ring oscillators.}
    \label{fig:ringosc}
\end{figure}

\section{Testchip Design and Measurement Results \label{sec:meas}}

We designed a testchip to assess strong PUF performance metrics, such as uniformity, uniqueness, and bit error rate (BER). Our objective is to evaluate the impact of $g$ in response stability, which will provide essential data to explore the security/stability trade-off in section \ref{sec:sec}.

\subsection{Testchip Design}

We designed a testchip in 65~nm CMOS technology. Our testchip includes 10 instances of NMQ-RO. The XOR-NMQ-RO composition is realized through post processing responses from these instances.

\begin{figure}[t]
    \centering
    \includegraphics[scale=1]{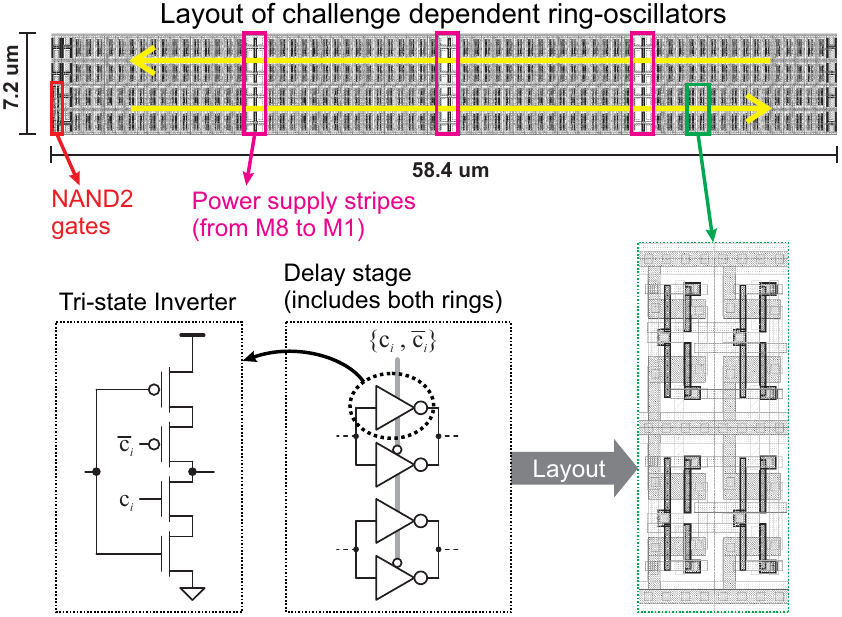}
    \caption{Layout and schematic of challenge dependent ring-oscillator.}
    \label{fig:laysch}
\end{figure}

The CMOS implementation of NMQ-RO uses custom designed ring-oscillators with an even number of challenge enabled tri-state inverters, and a NAND gate for control, as shown in Fig. \ref{fig:ringosc}. Layout of ring-oscillators, and the schematic for tri-state inverters are shown in Fig. \ref{fig:laysch}. No special circuit technique was used to compensate for temperature or noise.

\begin{figure}[t]
    \centering
    \includegraphics[scale=1]{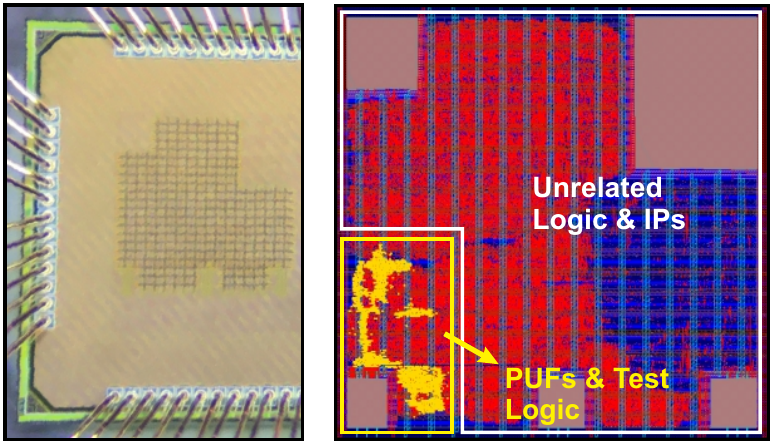}
    \caption{Die photo of fabricated chip in 65~nm CMOS process.}
    \label{fig:chip}
\end{figure}

We used automated synthesis, placement, and routing tools to design the trap counter and surrounding test logic, which are shown in Fig. \ref{fig:chip}, highlighted in yellow. An additional test counter was created to keep track of the total number of toggles, which is only used for experimental reasons.

\subsection{Measurement Results}

\begin{figure}[t]
    \centering
    \includegraphics[scale=1]{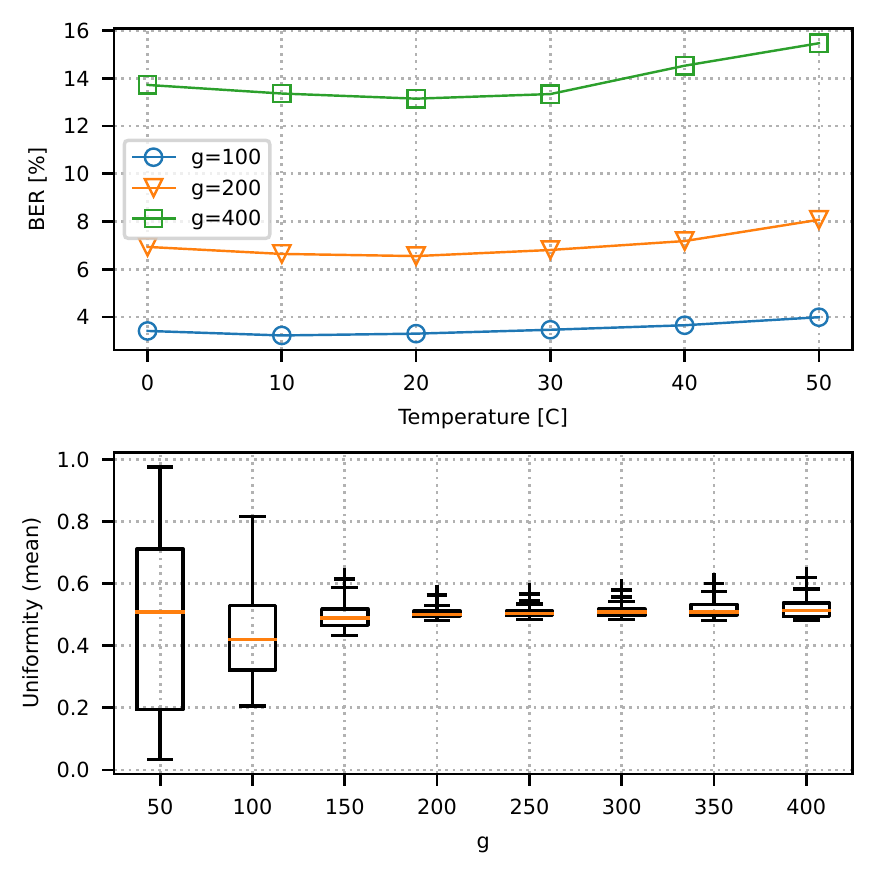}
    \caption{NMQ-RO (a) average BER, and (b) distribution of instance uniformity, for multiple values of $g$, using 10 k CRPs.}
    \label{fig:ber_lev}
\end{figure}

Measurements were performed on a total of 60 NMQ-RO instances, spread over 6 dies. Each die also included one APUF instance. We performed enrollment at \temp{20} with a single evaluation (no temporal majority voting). To calculate BER, CRPs are evaluated 100 times at each temperature from \temp{0} to \temp{50}. The reported metrics were calculated over all 60 instances of NMQ-RO.

Bit error rate measures the stability of PUF responses. BER of NMQ-RO depends on the parameter $g$. Fig. \ref{fig:ber_lev} (a) shows BER for different values of $g$, measured using 10 k CRPs. At \temp{20}, we measured BER of 3.3\%, 6.5\%, and 13.1\%, for g=100, g=200, and g=400, respectively. Other temperatures in the \temp{0} -- \temp{50} range show a small BER variation compared to their respective enrollment temperature.

\begin{figure}[t]
    \centering
    \includegraphics[scale=1]{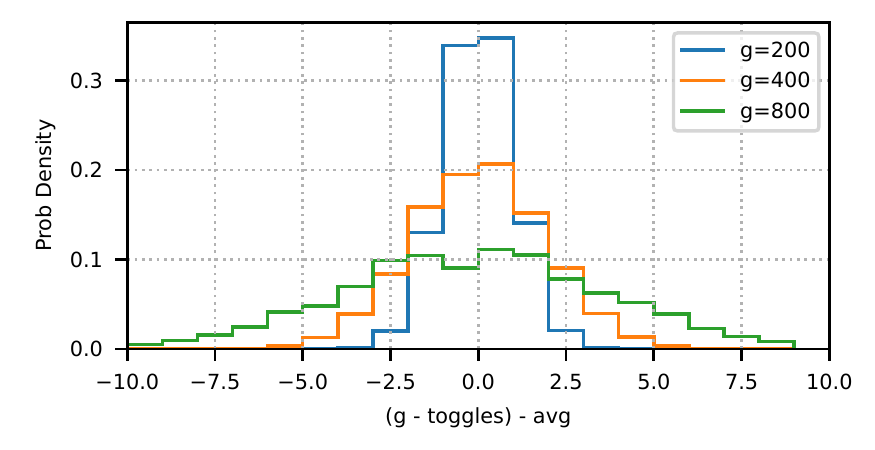}
    \caption{NMQ-RO histogram of trap counter final value minus number of toggles for various $g$, using 100 k challenges. The average value was subtracted to center histograms at zero.}
    \label{fig:histxy}
\end{figure}

\begin{figure}[t]
    \centering
    \includegraphics[scale=1]{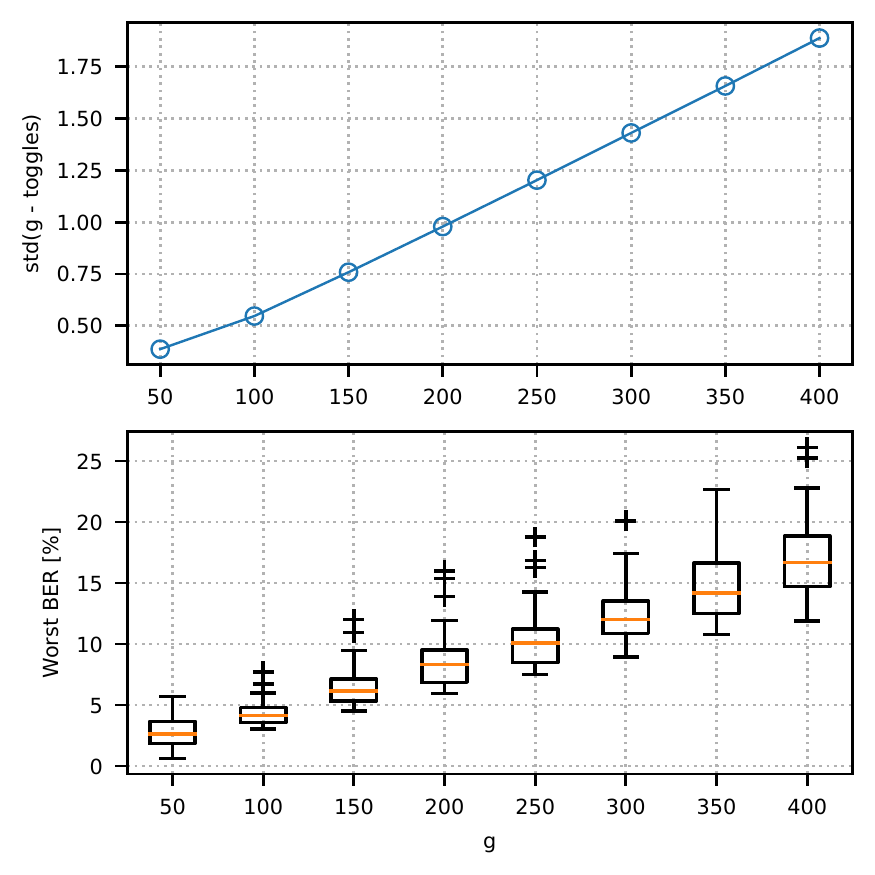}
    \caption{NMQ-RO (a) standard deviation of trap counter final value minus the number of toggles; and (b) worst BER distribution for all instances from \temp{0} to \temp{50}. Metrics calculated using 10 k CRPs.}
    \label{fig:std_ber}
\end{figure}

Fig. \ref{fig:ber_lev} (b) depicts the uniformity distribution (mean values) for all instances with respect to $g$ values. Uniformity was calculated using 10 k CRPs. The larger uniformity spread observed when $g$ is less than 200 is caused by performance mismatches between ROs. Hence, when NMQ-RO runs with small $g$ values, a significant number of challenges will not accumulate a sufficiently large delay difference to move $\flr{g D_p/D_q}$ away from its \textit{average} value, leading to biased responses.

The histogram of trap counter final value minus number of toggles is shown in Fig. \ref{fig:histxy}, for different values of $g$, using 100 k challenges. The average value of each distribution was subtracted to center them at zero. The histogram becomes wider as $g$ increases, as a result of more cycles available to accumulate the delay difference present in the two ROs. Similar results are also shown in Fig. \ref{fig:std_ber} (a), where standard deviation is plotted for a range of $g$ values, using 10 k CRPs. Fig. \ref{fig:std_ber} (b) plots the worst BER distribution for all instances, at the respective $g$, also using 10 k CRPs.

\begin{figure}[t]
    \centering
    \includegraphics[scale=1]{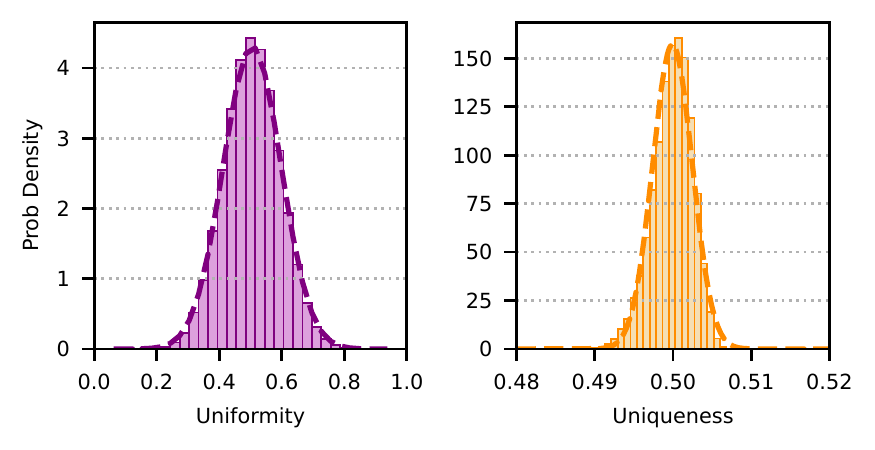}
    \caption{NMQ-RO (a) uniformity for 1M CRPs (all instances combined), and (b) uniqueness for 1M CRPs across all instances.}
    \label{fig:unif_uniq}
\end{figure}

The uniformity and uniqueness histograms were calculated using 1M CRPs per instance, and exhibits near ideal behavior. The uniformity histogram, for all instances combined, is shown in Fig. \ref{fig:unif_uniq} (a). Uniformity mean is 0.505 with 0.093 standard deviation, which represents that, on average, 50.5\% of responses are 0, and 49.5\% are 1. The uniqueness histogram is shown in Fig. \ref{fig:unif_uniq} (b). We used multi-bit responses of 32-bit. The uniqueness mean is 0.500 with 0.003 standard deviation, which indicates that, on average, 50.0\% of response bits were different in every group of 32 responses, for the same challenges, across multiple instances.

\begin{figure}[t]
    \centering
    \includegraphics[scale=1]{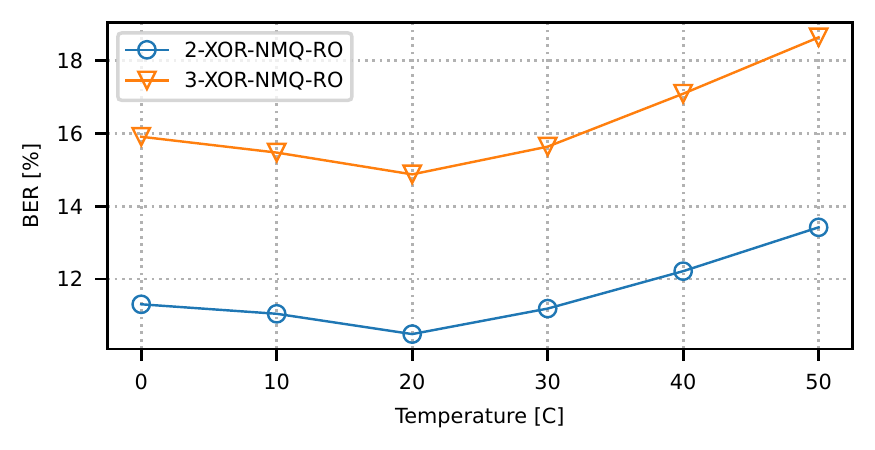}
    \caption{BER of 2-XOR-NMQ-RO, and 3-XOR-NMQ-RO compositions, using 5 k CRPs.}
    \label{fig:ber_xor}
\end{figure}

Our results report data for a total of 18 instances of 3-XOR-NMQ-RO, and 30 instances of 2-XOR-NMQ-RO. Fig. \ref{fig:ber_xor} shows mean BER calculated using 5 k challenges, from \temp{0} to \temp{50}. The lowest BER occurs at the enrollment temperature of \temp{20}, while the peak BER value is found at \temp{50}. The worst BER for 2-XOR-NMQ-RO, and 3-XOR-NMQ-RO is 13.4\%, and 18.6\%, respectively.

Better response stability is likely possible via the utilization of more specialized circuit techniques. Related works that seek to increase response stability are discussed in section \ref{sec:relworks}. Moreover, NMQ-PUFs may be implemented using circuit techniques other than from ring-oscillators, which could potentially produce PUFs with improved BER.

\section{Security Assessment \label{sec:sec}}

The goal of learning attacks is to find a representation of the PUF entropy source, such that it can be used to accurately predict responses to unseen challenges. In this section, we first develop the concept of uniqueness sensitivity to entropy source, showing how uniqueness sensitivity can provide rapid insight on PUF resilience to learning attacks. Next, we report extensive experimental results with various learning attack techniques.

\subsection{Uniqueness Sensitivity to Entropy Source}

\begin{figure*}[t]
    \centering
    \includegraphics[scale=1]{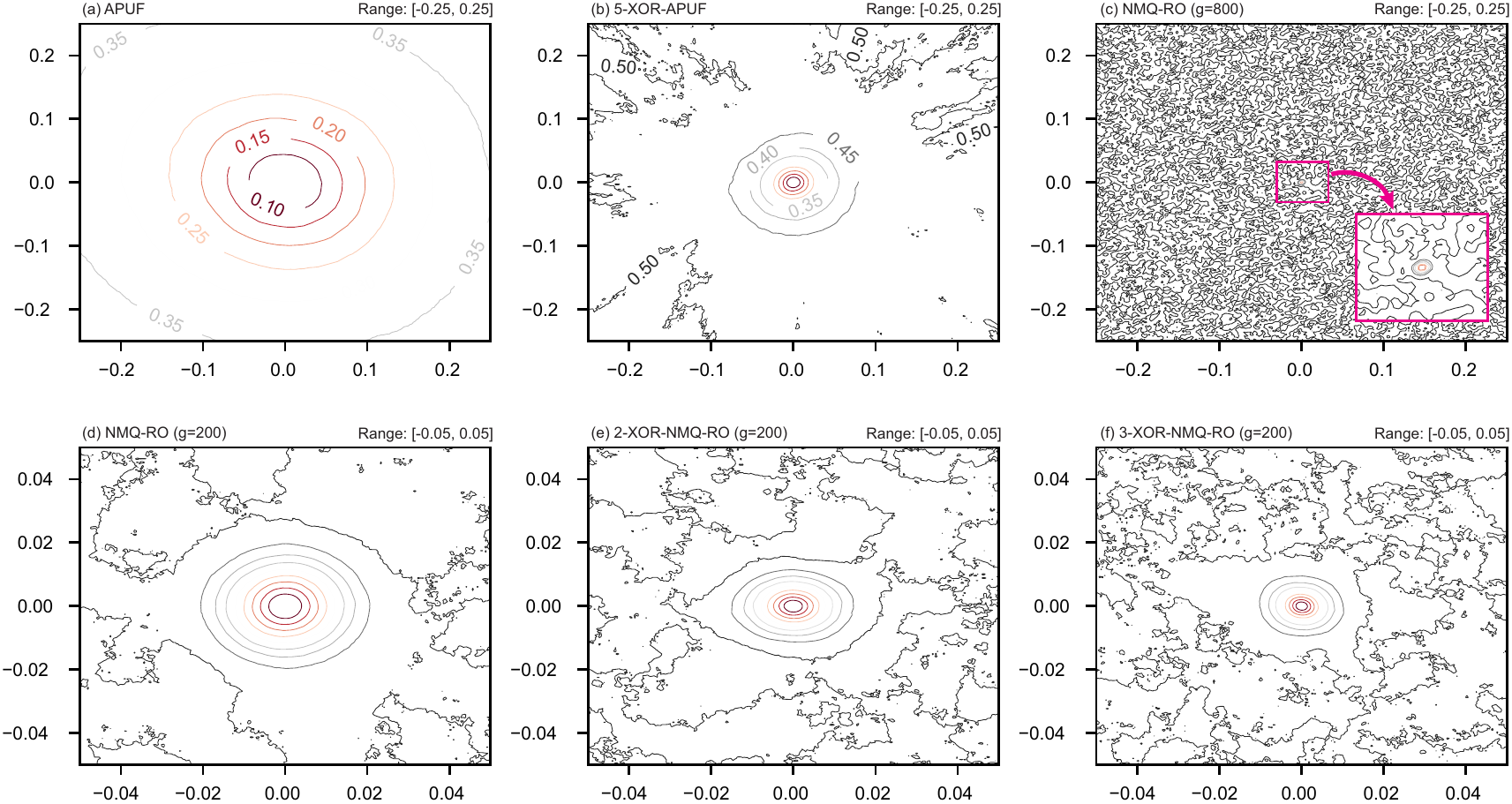}
    \caption{Contour plots of uniqueness surface exploration over entropy source space; (a) APUF, (b) 5-XOR-APUF; (c) NMQ-RO (g=800); (d) NMQ-RO (g=200); (e) 2-XOR-NMQ-RO (g=200); (f) 3-XOR-NMQ-RO (g=200).}
    \label{fig:cost}
\end{figure*}

We investigate a technique to gain rapid insight on learning resilience of strong PUFs by exploring the uniqueness metric, and its interaction with PUF entropy source. Experiments using this technique take less than one hour to complete, without any specialized hardware accelerators. This technique does not replace comprehensive attack experiments, but instead, offers valuable insight during design stage, allowing fast iteration between different PUF architectures, and design parameters.

The entropy source of PUFs originates at random deviations in device parameters that occur during IC manufacturing. Small deviations in the entropy source of a strong PUF should produce unique responses. If $\bm{\theta}$ denotes entropy source parameters, the uniqueness of two different PUF instances is given by $U(\bm{\theta}_0, \bm{\theta}_1)$.

\medbreak
\begin{definition}\label{def:uniq}
Uniqueness sensitivity is the rate of change in uniqueness with respect to the entropy source.
\end{definition}
\medbreak

The high-dimensional space of PUF parameters makes practical use of uniqueness sensitivity non-trivial. To visualize the impact of small entropy source changes in uniqueness, we use a curvature exploration method, described in \cite{othersLoss2018}. We pick two random direction vectors $\bm{\delta}$ and $\bm{\eta}$, of same dimension as $\bm{\theta}$, and evaluate

\begin{align} \label{eq:cost}
f(\alpha, \beta) = U(\bm{\theta}_0, \bm{\theta}_0 + \alpha \bm{\delta} + \beta \bm{\eta}).
\end{align}

Parameters $(\alpha, \beta)$ are scalars, which are swept as input coordinates to calculate uniqueness over a set of challenges. When $(\alpha=0, \beta=0)$, uniqueness will be evaluated for two identical instances (same entropy source), resulting in zero. As $(\alpha, \beta)$ move away from the origin,  the distance to the original entropy source, $\bm{\theta}_0$, will increase, hence the uniqueness between the original instance, and the new instance parameterized by $\bm{\theta}_0+\alpha\bm{\delta} + \beta\bm{\eta}$, will diverge from $0$ and center around $0.5$.

Fig. \ref{fig:cost} shows contour plots for uniqueness in different PUF architectures. In (a), (b), and (c) the input coordinate range to Eq. \ref{eq:cost} is $\alpha \in [-0.25, 0.25]$, and $\beta \in [-0.25, 0.25]$. In (a), the APUF uniqueness is zero at $(0,0)$, and gradually increases as $\alpha$ and $\beta$ move away from the origin. Similar behavior is observed in (b), where the 5-XOR-APUF uniqueness is zero at $(0,0)$, and increases as we move away from the origin. However, the 5-XOR-APUF shows a much steeper uniqueness increase, with the ideal contours of 0.5 appearing closer to the origin. 

The uniqueness contour plots for NMQ-RO and XOR-NMQ-RO are plotted in Fig. \ref{fig:cost} (c), (d), (e), and (f). In (c), the NMQ-RO (g=800) is plotted with the same range of input coordinates as (a), and (b). Uniqueness sensitivity is such that contours with values smaller than 0.5 are barely visible, requiring a zoomed graph inset.

For plots in Fig. \ref{fig:cost} (d), (e), and (f), the input coordinate range is 5x narrower, $\alpha \in [-0.05, 0.05]$, and $\beta \in [-0.05, 0.05]$. The contour plots for NMQ-RO (g=200) is shown in (d). Extra instances of NMQ-RO (g=200) are XORed to form 2-XOR-NMQ-RO in (e) and 3-XOR-NMQ-RO in (f). We observed that XORing additional NMQ-RO instances caused a small increase in uniqueness sensitivity, and removed noticeable patterns in the contour lines near the origin.

\medbreak
\begin{observation}\label{obs:uniqml}
High uniqueness sensitivity to entropy source deviations is an indicator of strong PUF robustness to learning attacks.
\end{observation}
\medbreak

\begin{table*}[t]
    \centering
    \caption{Learning attack results.}
    \label{tab:attacks}
    \begin{threeparttable}
        \begin{tabular}{@{}llrrrlllrl@{}}
\toprule
                                                                               & \textbf{Quant.}           & \multicolumn{1}{l}{\textbf{Chains}} & \multicolumn{1}{l}{\textbf{Worst BER}} & \textbf{CRPs} & \textbf{Database Type} & \textbf{Attack Type} & \multicolumn{1}{l}{\textbf{Accuracy}} & \textbf{Time} \\ \midrule
APUF                                                                           & Typ                       & 1                                   & 3\%                                 & 10k           & Model \& Silicon               & LR        & 99.8\%   & 1 s       \\ \midrule
\multirow{2}{*}{XOR-APUF}                                                      & \multirow{2}{*}{Typ}      & 8                                   & \multirow{2}{*}{20\%}               & 10k           & Silicon \cite{attackRelb2015}  & CMA-ES    & 88.0\%   & 1.6 min   \\
                                                                               &                           & 8                                   &                                     & 1M            & Model                          & DNN       & 94.0\%   & 9 min     \\ \midrule
\multirow{3}{*}{\begin{tabular}[c]{@{}l@{}}NMQ-RO\\ (g=200)\end{tabular}}      & \multirow{3}{*}{NMQ}      & \multirow{3}{*}{1}                  & \multirow{3}{*}{8\%}                & 1M            & Model \& Silicon               & LR        & 50.1\%   & 10 s      \\
                                                                               &                           &                                     &                                     & 1M            & Model \& Silicon               & Fourier   & 50.0\%   & 1 h       \\
                                                                               &                           &                                     &                                     & 1M            & Model \& Silicon               & DNN       & 95.1\%   & 12 h      \\ \midrule
\multirow{2}{*}{\begin{tabular}[c]{@{}l@{}}NMQ-RO\\ (g=400)\end{tabular}}      & \multirow{2}{*}{NMQ}      & \multirow{2}{*}{1}                  & \multirow{2}{*}{15\%}               & 1M            & Model \& Silicon               & DNN       & 90.0\%   & 12h       \\
                                                                               &                           &                                     &                                     & 5M            & Model \& Silicon               & DNN       & 91.4\%   & 2.5 days  \\ \midrule
\multirow{2}{*}{\begin{tabular}[c]{@{}l@{}}NMQ-RO\\ (g=800)\end{tabular}}      & \multirow{2}{*}{NMQ}      & \multirow{2}{*}{1}                  & \multirow{2}{*}{29.4\%}             & 1M            & Model \& Silicon               & DNN       & 75.0\%   & 12 h      \\
                                                                               &                           &                                     &                                     & 5M            & Model \& Silicon               & DNN       & 86.5\%   & 2.5 days  \\ \midrule
\begin{tabular}[c]{@{}l@{}}NMQ-RO\\ (g=5000)\end{tabular}                      & NMQ                       & 1                                   & *                                   & 50M           & Model                          & DNN       & 50.7\%   & 4 days    \\ \midrule
\multirow{4}{*}{\begin{tabular}[c]{@{}l@{}}XOR-NMQ-RO\\ (g=200)\end{tabular}}  & \multirow{4}{*}{NMQ}      & \multirow{2}{*}{2}                  & \multirow{2}{*}{13.4\%}             & 20M           & Model                          & DNN       & 50.3\%   & 7 days    \\
                                                                               &                           &                                     &                                     & 10M           & Silicon                        & DNN       & 49.6\%   & 3.5 days  \\
                                                                               &                           & \multirow{2}{*}{3}                  & \multirow{2}{*}{18.6\%}             & 20M           & Model                          & DNN       & 49.9\%   & 7 days    \\
                                                                               &                           &                                     &                                     & 10M           & Silicon                        & DNN       & 50.2\%   & 3.5 days  \\ \bottomrule
\end{tabular}

        \begin{tablenotes}[para,flushleft]
            Notes: BER is the worst bit error rate from \temp{0} to \temp{50}, with challenges enrolled at \temp{20}. Worst BER and accuracy results for XOR-APUF under CMA-ES attack as reported in \cite{attackRelb2015} using a 2-XOR-4-way construction (similar to 8-XOR-APUF).
        \end{tablenotes}
    \end{threeparttable}
\end{table*}

Typically, learning attacks use differentiable neural network architectures trained by stochastic, gradient descent algorithms. Those optimization algorithms search for weights and biases which \textit{minimize} a loss function between the collected CRP database, and network predictions. If the learned network parameters are a representation of the strong PUF entropy source, we can argue that the contour plots of uniqueness over the entropy source space, shown in Fig. \ref{fig:cost}, capture the impact of a particular PUF architecture on the loss function used for training. Therefore, our results suggest that strong PUFs with high uniqueness sensitivity will likely produce harder to optimize loss functions.

\subsection{Learning Attack Results}

We performed experiments to assess the security of NMQ-PUFs. We used CRP databases generated from a high-level model, and from our silicon implementation. Both the high-level model and the CRP database collected from silicon are publicly available \cite{resCode, resDataset}. Our findings are summarized in Table \ref{tab:attacks}. The \textit{Quant.} column defines the quantization method used, either non-monotonic, or typical. The \textit{Chains} column is the number of XORed instances, or $k$. The \textit{Worst BER} represents the worst case measured BER over \temp{0} to \temp{50} temperature range. The \textit{Database Type} column specifies the origin of CRPs used in the attack -- \textit{Model} refers to the high-level model, while \textit{Silicon} refers to CRPs from our testchip. If both database types are specified, it indicates the experiment was run twice, once with each database, obtaining very similar results for both runs.

The APUF, shown in the first line of Table \ref{tab:attacks}, uses typical quantization and has worst BER of 3\%. An attack using logistic regression (LR) was able to achieve 99.8\% accuracy. Training used 10k CRPs and was completed in one second. We also performed attacks on the 8-XOR-APUF, which uses an XOR to combine 8 APUFs. The overall composition has 20\% BER. In \cite{attackRelb2015}, authors describe a very effective technique using evolution strategies (ES) for attacking XOR-APUFs. They implemented a covariance matrix adaptation (CMA-ES) algorithm that uses response stability as side-channel information. The key insight is that, in typically quantized PUFs, CRPs with small delay difference are more susceptible to noise. Authors demonstrated that increasing the number of APUFs is ineffective to protect against CMA-ES. They were able to model an APUF construction with 8 XOR operations with 88.0\% accuracy using CMA-ES. Training took 1.6 min and used 10 k CRPs.

Deep-neural networks (DNNs) are emerging as an efficient attack technique capable of learning complex PUF structures. We use a 12-layer DNN architecture proposed in \cite{attackDNN2019} for all our DNN attacks. The input and output layers have 64, and 2 units, respectively. Hidden layers have 2000 units. Table \ref{tab:attacks} shows the DNN model was able to obtain 94.0\% accuracy when learning an 8-XOR-APUF. The training of the DNN model used 1 M CRPs and took 9 min. Unlike CMA-ES, the DNN does not use response stability as side-channel information.

The learning resilience of NMQ-PUFs was first assessed using LR. We used 1M CRPs to train an LR model with NMQ-RO responses using g=200, which resulted in 8\% BER. The modeling accuracy obtained was 50.1\%, which supports Obs. \ref{obs:linsep}, suggesting that NMQ-PUF responses can not be learned with linear models.

Many attack techniques in the literature assume typically quantized responses. Such techniques will likely fail at predicting NMQ responses. For example, CMA-ES assumptions on noisy CRPs do not hold for NMQ-PUF -- small performance differences are not correlated to response stability. Similarly, other attack techniques that rely on a limited challenge space are unlikely to model NMQ-RO responses. As mentioned before, NMQ-RO uses challenge dependent ROs, which may be seen as a pool of $2^{64}$ different ROs. For example, in \cite{attackROPac2015}, the polynomial-size decision list cannot represent our architecture. Another similar attack using Fourier analysis was introduced in \cite{attackROFourier2018}, and as shown in Table \ref{tab:attacks}, it was not able to obtain generalized learning of NMQ-RO.

The NMQ-RO (g=200) demonstrated substantial improvement in learning resilience compared to APUF. Nevertheless, Table \ref{tab:attacks} shows that DNN attacks are capable of learning NMQ-ROs using g=200, g=400, and g=800. The trade-off between final trap counter value and stability is present, where larger $g$ values increase learning resilience, but decrease response stability. An exploratory DNN attack was performed with g=5000. The DNN model was not able to model NMQ-RO using g=5000, which shows that increasing $g$ will, eventually, make the PUF resistant to this attack -- aside from the significant reduction in response stability that makes such configuration unpractical.

To find a compromise between learning resilience and response stability, we explored compositions that use NMQ-RO. Since compositions tend to decrease response stability, we use NMQ-RO at a moderate $g$ value (g=200). Results in Table \ref{tab:attacks} show the DNN attack was not able to obtain generalized learning of XOR-NMQ-RO. Moreover, both implementations, with 2, and 3 chains, show acceptable BER of 13.4\% and 18.6\%, while preserving the PUF entropy source parameters from DNN attacks using up to 20M CRPs.

\section{Conclusion\label{sec:conc}}

We introduced a non-monotonic quantization technique to increase security of strong PUF architectures. We implemented a non-monotonically quantized ring-oscillator based strong PUF in 65~nm CMOS technology. Our experiments show that the resulting PUF has increased security against learning attacks. Measurement results also show the proposed PUF has less than 13.4\% BER over a temperature range of \temp{0} to \temp{50}. Moreover, we introduced the concept of uniqueness sensitivity to entropy source, showing how it can provide rapid insight on PUF resilience to learning attacks.

\bibliographystyle{plain}
\bibliography{nmq}

\vspace{-0.4in}
\begin{IEEEbiography}[{\includegraphics[width=1in,height=1.25in,clip,keepaspectratio]{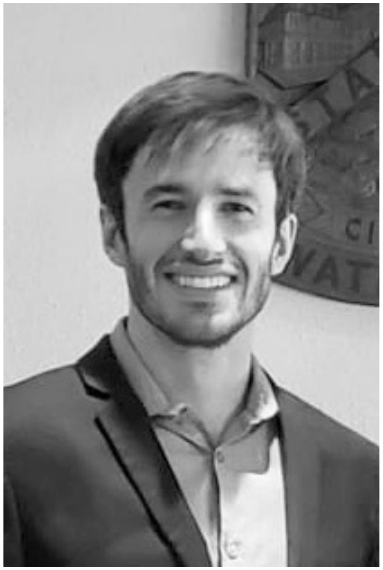}}]{Kleber Stangherlin}
Kleber received his B.Sc. in Electrical Engineering at PUCRS, and M.Sc. in Microelectronics at UFRGS, both in Brazil. He has more than 6 years of industry experience designing security focused integrated circuits. He had key contributions to the cryptographic cores and countermeasures used in the first EAL 4+ certified chip designed in the southern hemisphere. Currently, Kleber is pursuing a PhD at University of Waterloo in Canada, where he conducts research in hardware security.
\end{IEEEbiography}

\vspace{-0.4in}
\begin{IEEEbiography}[{\includegraphics[width=1in,height=1.25in,clip,keepaspectratio]{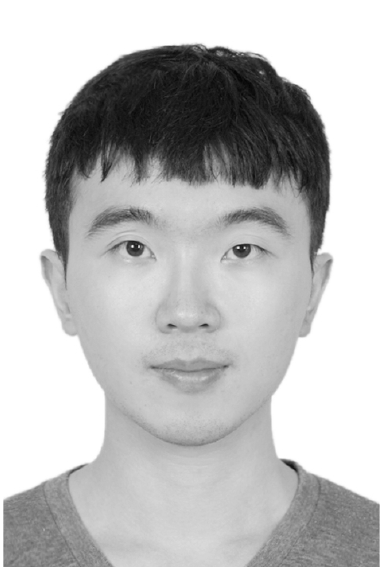}}]{Zhuanhao Wu}
Zhuanhao Wu (S'19) received the B.S. degree in computer science and technology from Nankai University, Tianjin, China in 2017, and the MASc degree in electrical and computer engineering from University of Waterloo, Ontario, Canada, in 2019, where he is currently pursuing the Ph.D. degree with the department of Electrical and Computer Engineering. His current research interests include hardware security, machine learning, and computer architecture.
\end{IEEEbiography}

\vspace{-0.4in}
\begin{IEEEbiography}[{\includegraphics[width=1in,height=1.25in,clip,keepaspectratio]{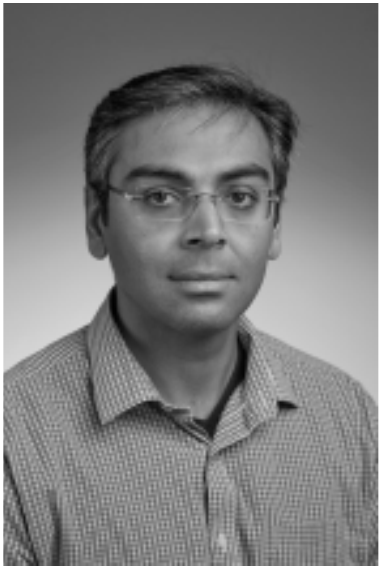}}]{Hiren Patel}
Hiren Patel is a Professor in the Department of Electrical and Computer Engineering at the University of Waterloo. Prior to the University of Waterloo, Hiren was a postdoctoral fellow at the University of California, Berkeley working in the Ptolemy group with Edward A. Lee. His research is in the design, analysis, and implementation of computer hardware and software. Currently, his research areas of interest are in real-time embedded systems, computer architecture, hardware architectures for machine learning and artificial intelligence, and security.
\end{IEEEbiography}


\vspace{-0.4in}
\begin{IEEEbiography}[{\includegraphics[width=1in,height=1.25in,clip,keepaspectratio]{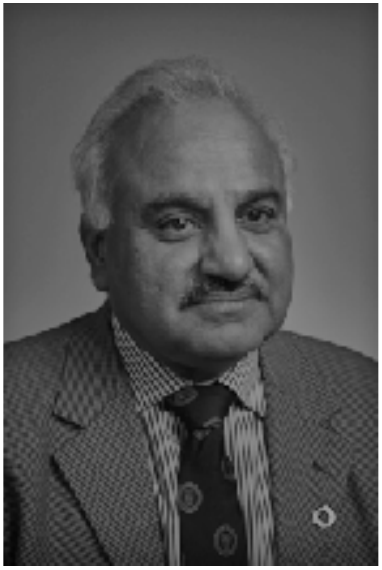}}]{Manoj Sachdev}
Manoj Sachdev is a Professor in the Department of Electrical and Computer Engineering at the University of Waterloo. He has contributed to over 225 conference and journal publications, and has written 5 books. He also holds more than 30 granted US patents. Along with his students and colleagues, he has received several international research awards. He is a Fellow of the Institute of Electrical and Electronics Engineers (IEEE), and Fellow of the Engineering Institute of Canada. Professor Sachdev serves on the editorial board of the Journal of Electronic Testing: Theory and Applications. He is also a member of program committee of IEEE Design and Test in Europe conference.
\end{IEEEbiography}

\vfill

\end{document}